\documentclass{aa}
\usepackage{psfig}
\usepackage{epsfig}

\def \hcm {\hbox {\ifmmode $ cm$^{-2}\else cm$^{-2}$\fi}}

\def \arcsec {\hbox{$^{\prime\prime}$}}

\def\approxgt{\mathrel{\hbox{\rlap{\lower.55ex \hbox {$\sim$}}
        \kern-.3em \raise.4ex \hbox{$>$}}}}
\def\approxlt{\mathrel{\hbox{\rlap{\lower.55ex \hbox {$\sim$}}
        \kern-.3em \raise.4ex \hbox{$<$}}}}

\begin{document}
%----------------------------------

\title{X-ray flares reveal mass and angular momentum of the Galactic Center black hole}

\author{B. Aschenbach\inst{1} 
        \and N. Grosso\inst{2}
          \and D. Porquet\inst{1}         
            \and P. Predehl\inst{1}
       }
\offprints{Bernd Aschenbach\\ (bra@mpe.mpg.de)}

\institute{Max-Planck-Institut f\"{u}r extraterrestrische Physik,
P.O. Box 1312, Garching bei M\"{u}nchen D-85741, Germany
\and Laboratoire d'Astrophysique de Grenoble, Universit\'e Joseph-Fourier, 
BP53, 38041 Grenoble Cedex 9, France
}
\date{Received December 17, 2003 ; Accepted January 23, 2004  }

\abstract{We have analysed the light curve of the two brightest X-ray flares
 from the Galactic Center black 
hole, one flare observed by {\sl{XMM-Newton}} on October 3, 2002 (Porquet et al. \cite{P2003}), 
and the other flare observed by  {\sl Chandra} on October 26, 2000 (Baganoff et al. \cite{B2001}). 
The power density spectra show five  
 distinct peaks at periods of $\sim$ 100\,s, 219\,s, 700\,s, 1150\,s,
 and 2250\,s common to both observations 
within their estimated measurement uncertainties. The power density 
spectrum of the recently reported infrared flare of June 16, 2003 (Genzel et al. \cite{Ge2003}) 
shows distinct peaks at two, if not three, periods (including the 1008$\pm$120\,s  
infrared period), which are consistent with the X-ray periods. The remaining two periods 
could not be covered by the infrared measurements. 
Each period can be identified with one of the characteristic 
gravitational cyclic modes associated with accretion disks, i.e. either Lense-Thirring 
precession, Kepler orbital motion and the vertical and radial epicyclic oscillation modes, 
in such a way that a consistent value for the black hole mass of  
M$\sb{\rm{BH}}$ = 2.72$\sp{+0.12}\sb{-0.19}$ $\times$ 10$\sp 6$M$_\odot$
and the angular momentum $a$ = 0.9939$\sp{+0.0026}\sb{-0.0074}$ is obtained. 
The available data on M$\sb{\rm{BH}}$
 derived from studies of the orbital motion of the S2 (S0-2) star
(Sch\"odel et al. \cite{S2002}, Ghez et al. \cite{Gh2003}) agree with our findings. 
Finally we discuss some implications 
of the fairly high value for the angular momentum.

\keywords{Galaxy: center -- X-rays: individuals: Sgr\,A* -- X-rays: general -- Radiation mechanisms: general }
}
\titlerunning{Mass and angular momentum of the GC black hole}
\authorrunning{Aschenbach et al.}
\maketitle

\section{Introduction}

The region towards the Galactic Center (GC) has been recently resolved in X-rays by {\sl Chandra}
measurements (Muno et al. \cite{MBB2003}, Baganoff et al. \cite{BMM2003}). 
A source was identified as X-ray counterpart 
of the GC black hole Sgr\,A*. The quiescent X-ray luminosity turns out to be extraordinary 
low (2.2$\times$10$^{33}$\,erg\,s$^{-1}$, Baganoff et al. \cite{BMM2003}) for a mass accreting black hole, 
given its mass which has been measured through the motion of the S0-2 star orbiting the black hole. 
Mass values range from 2.2$\times$10$\sp 6$M$_\odot$
 (lower limit of Sch\"odel et al. \cite{S2002}), 3.6 $\times$
10$\sp 6$M$_\odot$ (Genzel et al. \cite{Ge2003}) to 
 4.1$\pm$0.6 $\times$10$\sp 6$M$_\odot$ for a distance 
of 8 kpc (Ghez et al. \cite{Gh2003}). Various models to explain the 
low luminosity have been proposed and discussed quite 
controversially including low accretion rates  and radiatively inefficient accretion flows, 
for example. For a brief overview see for instance Melia \& Falcke ({\cite{MF2001}), 
Baganoff et al. (\cite{B2001}), Porquet et al. (\cite{P2003}) 
and Yuan et al. (\cite{YQN2003}). The quiescent state 
is interrupted by occasional, sometimes very bright, flares. The first bright flare with a peak luminosity of 
1$\times$ 10$^{35}$\,erg\,s$^{-1}$ (Baganoff et al. \cite{B2001}) was observed with the {\sl Chandra}
ACIS-I instrument on October 26, 2000. An even brighter X-ray flare with a peak luminosity of 
3.6$\times$10$^{35}$\,erg\,s$^{-1}$ occurred on October 3, 2002 and was caught with the 
{\sl{XMM-Newton}} EPIC instruments (Porquet et al. \cite{P2003}). It is generally accepted that the 
radiation is from or at least channeled through the accretion disk orbiting the central black hole.      
Some models for the creation of flares have been brought forward, e.g., Markoff et al. ({\cite{M2001}) suggest 
extra electron heating near the black hole, Liu \& Melia (\cite{LM2002}) propose 
a sudden enhancement of accretion and 
Nayakshin \&  Sunyaev (\cite{NS2003}) prefer 
stars passing the disk. We think that the high luminosity involved in the flares 
suggests that somehow the inner regions of the accretion disk play a major role, basically 
because of the increased gravitational potential. Sporadic events like flares are then likely 
to take us closer to the black hole and enable us to discover more easily periodic or quasi-periodic 
changes in the light curve of a flare, like the Kepler frequency of the innermost marginally stable 
orbit, and others, by which mass and angular momentum of the central black hole could be 
determined. That this idea might have some bearing has been recently demonstrated by the discovery
 of a 16.8$\pm$2 min period in two infrared flares (Genzel et al. \cite{Ge2003}). Genzel et al. tentatively 
assign the period to the length of the marginally stable orbit and, adopting a mass of 3.6 $\times$
10$\sp 6$M$_\odot$, they calculate an angular momentum of $a$ = 0.5 for the black hole. 

\par

In section 2 we show the light curves of the two brightest X-flares, which we 
have analysed including but separated from the time sections before and after the flare proper, as 
well as two very long (11 hours and 46 hours) X-ray light curves representing the so-called 
quiescent state. Like the two flares one event was observed by {\sl{XMM-Newton}} and the other event 
was covered by {\sl{Chandra}}. These two observations were primarily analysed 
for cross-checking with the flare characteristics. Section 2 also contains the power density 
spectra (PDS) of the relevant light curves. Five groups of characteristic frequencies with at least 
two detections per group have been found. In section 3 
we describe the relation of these periods with the cyclic gravitational modes thought 
to be possibly created   
in accretion disks, i.e. the Lense-Thirring precession frequency, the Kepler frequency and 
the vertical and radial epicyclic frequencies. In section 3 we also describe how we derive the 
values for M$\sb{\rm{BH}}$ and the angular momentum $a$
 (the Kerr parameter) from the frequencies measured. 
We include in our analysis the power density spectra of the two infrared flares published 
by Genzel et al. (\cite{Ge2003}). 
Finally we discuss a few implications which result 
from the fairly high value of $a$ we have obtained. 
  
\section{Observations and analysis}

\subsection{Light curves}

Figure~\ref{fig1}a shows the light curve of the October 3, 2002 flare  observed by 
{\sl{XMM-Newton}} and published by Porquet et al. (\cite{P2003}) with a slightly different 
time binning. Figure~\ref{fig1}b shows the light curve of an $\sim$11 hours quiescent  
period obtained by {\sl{XMM-Newton}} on February 26, 2002. Figure~\ref{fig2}a 
displays the light curve of the very first flare reported from Sgr\,A* dated October 26, 2000,  
observed with {\sl Chandra} and published by Baganoff et al. (\cite{B2001}). We choose 
a slightly different energy band and a different binning. Sgr\,A* was observed for   
46.5 hours on May 5/6, 2002 without a bright flare occurring, and the {\sl Chandra}
light curve is shown in Figure~\ref{fig2}b. Because of the numerous 
peaks the source is apparently not in a true quiescent state but exhibits 
quite a number of smaller flux increases or small flares. 
The essential observational details are summarized in Table~\ref{obslog}.  
The {\sl{XMM-Newton}} data have been proprietary PI data, which by now are 
publicly available. The {\sl Chandra} data have been extracted from the public 
{\sl Chandra} archive (level 2 processed event list, provided by the {\sl Chandra X-ray Center}). 
We restrict our analyses to just these four data sets because 
they cover the two brightest flares and the other two observations have the longest 
exposure of the so-called quiescent state.
 For the timing analysis we added PN, MOS1 and MOS2 data in one 
set using counts, which were for {\sl XMM-Newton} extracted from a circle 
of 10\arcsec~radius with energies between 2.6 - 10 keV. 
For the {\sl Chandra} data the extraction radius was 1.5\arcsec~around the
archived source position, and the energy  band is  2.0 -- 8 keV.

\begin{table}
   \caption[]{Summary of observations analysed (c.f. Figs.~\ref{fig1},\ref{fig2}). 
                    Instr. I is EPIC PN+MOS1+MOS2 on {\sl XMM-Newton}, 
                    and instr. II is ACIS-I on {\sl Chandra}. 
                    Date is in day/month/year.}
    \label{obslog}
\begin{flushleft}
\begin{tabular}{lccccc}
            \noalign{\smallskip}
\hline
Instr.&observ.&total&flare&preced.& sampl.\\
         &          &time& time&time&time  \\
         &date       &        (min)    &  (min)     & (min)&(s)\\
\hline
\noalign {\smallskip}
I&03/10/02 &       254.1   &  56.33     & 197.8     &  2.6     \\
\noalign {\smallskip}
I&26/02/02 &           667.2   &  --     & --     &  2.6     \\
\noalign {\smallskip}
II&26/10/00 &          589.3   &  115.3     & 372.4     &  3.241     \\
\noalign {\smallskip}
II&25/05/02 &          2804.2   &  --     & --     &  3.141     \\
\hline
\end{tabular}
\end{flushleft}
%
%$\sp 1$The data during the time taken by the flare have been removed and replaced by a sequence
%of data Poissonian distributed around the mean level determined from the precursor data.
\end{table}

Because of the relatively broad point spread function of {\sl XMM-Newton} the extraction 
radius of 10\arcsec~contains emission from more objects than just Sgr\,A*. 
The {\sl Chandra} image reveals another four or five additional sources as well as diffuse 
emission. Baganoff et al. (\cite{BMM2003}) estimate a net count rate 
of 5.74$\pm$0.40 $\times$ 10$\sp{-3}$ 
count\,s$\sp{-1}$ for Sgr\,A* in its quiescent state. The power law spectrum has a photon index of 
$\Gamma\sb{\rm q}$ = 2.7$\sp{+1.3}\sb{-0.9}$ and an absorbing column density 
N$\sb{\rm{H,q}}$ =9.8$\sp{+4.4}\sb{-3.0} \times$ 10$\sp{22}$ cm$\sp{-2}$. 
Using the best fit values we estimate  a Sgr\,A* quiescent state count rate 
of 0.029 $\pm$ 0.003 count\,s$^{-1}$ for MOS\,1+MOS\,2+PN, so 
that the conversion factor is 5 $\pm$ 0.5 between the two instrument combinations. 
The mean EPIC count rate of Sgr\,A* in  its quiescent state 
amounts to 0.16 count\,s$\sp {-1}$ (c.f. Fig.~\ref{fig1}b), which means that the background sources are likely 
to contribute on average 0.013 count\,s$\sp {-1}$ or 82\% \ of the total during the Sgr\,A* 
quiescent state, provided that neither the background sources nor Sgr\,A* have changed 
significantly compared to the October 26, 2000 observation. 
When we subtract the count rate of the background from the flare 
observation (fig.~\ref{fig1}a) the peak count rate during the time preceding the actual flare 
has a level of 
0.125$\pm$0.03 count\,s$^{-1}$, which means 
that the precursor section is $\sim$4 times brighter than  Sgr\,A* in its quiescent state. It is interesting 
that flares observed in the infrared by Genzel et al. (\cite{Ge2003}) show about the same ratio 
of the flare peak intensity and the quiescent level. This suggests that the X-ray precursor and the infrared 
flares originated in the same process and that, given the huge X-ray flares, much brighter infrared 
flares still await their detections.

Conversion of instrument count rates to flux density requires knowledge of the source spectrum and 
column density of the photoelectrical absorption. This is available for the quiescent state measured 
by {\it{Chandra}} (Baganoff et al. \cite{BMM2003}), the October 26, 2000 flare measured by {\it{Chandra}} 
(Baganoff et al. \cite{B2001}) and the October 3, 2002 flare measured by {\it{XMM-Newton}} 
(Porquet et al. \cite{P2003}). Since both the spectrum and the absorption column density differ among 
the three observations we provide conversion factors for the EPIC count rate and the ACIS-I 
count rate for each of the three states. These are:
9.5 $\mu$Jy (EPIC counts/s)$\sp{-1}$, 47.6 $\mu$Jy (ACIS-I counts/s)$\sp{-1}$ (quiescent);
0.5$\sp{+1.1}\sb{-0.3}~\mu$Jy (EPIC counts/s)$\sp{-1}$, 2.8$\sp{+6.1}\sb{-2.0}~\mu$Jy (ACIS-I counts/s)$\sp{-1}$ 
({\it{Chandra}} flare);
12.3 $\mu$Jy (EPIC counts/s)$\sp{-1}$, 71.8 $\mu$Jy (ACIS-I counts/s)$\sp{-1}$ ({\it{XMM-Newton}} flare). 
The relatively large uncertainty of the conversion factor for the {\it{Chandra}} flare are due to the uncertain 
spectrum and column density. The energy flux densities are given for E = 1 keV.

\begin{figure}[t]
\psfig{figure=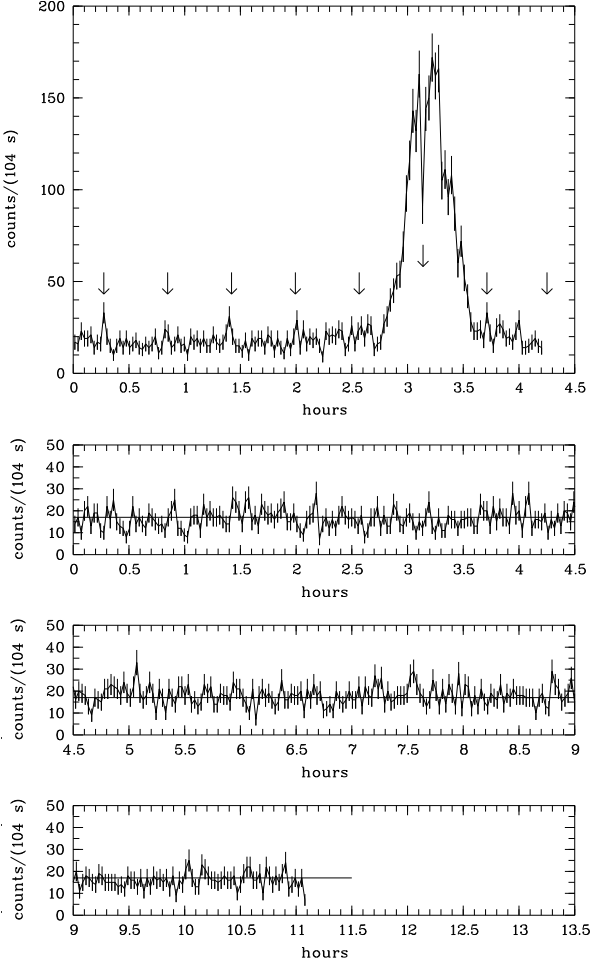,width=8cm,angle=0,%
bbllx=70pt,bblly=70pt,bburx=535pt,bbury=815pt,clip=}
%\vspace*{-0.56cm}%
\caption{EPIC light curves (MOS\,1+MOS\,2+PN) of the 
{\it{XMM-Newton}} observation 
of October 3, 2002 (upper panel, Fig. 1a) and 
the February 26, 2002 observation (lower three panels, Fig. 1b). 
Error bars indicate 1$\sigma$ uncertainties.
The horizontal line in the lower three panels corresponds to the mean count rate level. 
Arrows mark peaks associated with a 2178\,s periodic signal.
\vspace*{-0.4cm}
}
\label{fig1}
\end{figure}

\begin{figure}[t]
\psfig{figure=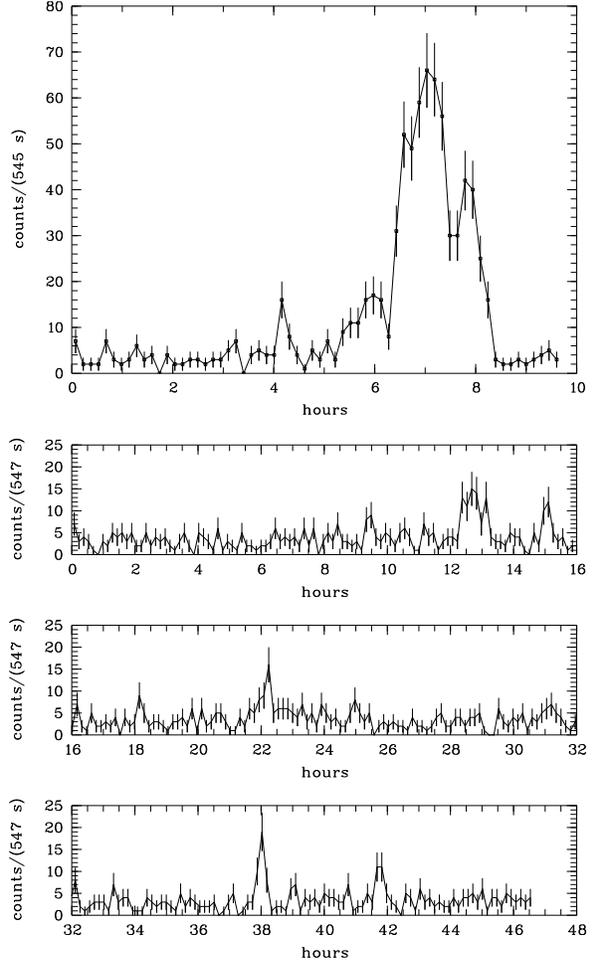,width=8cm,angle=0,%
bbllx=70pt,bblly=70pt,bburx=535pt,bbury=815pt,clip=}
%\vspace*{-0.56cm}%
\caption{ACIS-I light curves of the 
{\it{Chandra}} observation of October 26, 2000 (upper panel, Fig. 2a) and
the May 25, 2002 observation (lower three panels, Fig. 2b). Error bars indicate 1$\sigma$ uncertainties.
\vspace*{-0.4cm}
}
\label{fig2}
\end{figure}

\subsection {Power spectra}

From the light curves shown in Figures~\ref{fig1} and \ref{fig2},
 power density spectra (PDS) have been created
by a Fourier analysis of six data sets. Two sets cover the time section of just the flare, one each for
 {\sl XMM-Newton} (Fig.~\ref{fig3}a) and {\sl Chandra} (Fig.~\ref{fig4}a); 
another two sets cover the time sections before and after the
flare (Figs.~\ref{fig3}b and ~\ref{fig4}b). For these PDS the data of the flare 
proper have been removed and replaced by data with a mean flux
identical to that of the time preceding the flare assuming a Poissonian statistical distribution.
The final two sets correspond to the two observations of the quiescent level (Figs.~\ref{fig5} and \ref{fig6}).
We define the power spectral density psd$\sb{n}$ at a frequency f$\sb{n}$ and wave number n as
psd$\sb{n}$ = (a$\sb{n}\sp 2$+b$\sb{n}\sp 2$)/(2$\times\Delta t\sp 2)\times T$; a$\sb{n}$ and b$\sb{n}$ are the
Fourier coefficients, $\Delta t$ is the sampling or binning time and $T$ is the 
total observing time (Table~\ref{obslog}). 
The Fourier coefficients are in units of ACIS-I count\,s$^{-1}$ 
for the {\sl Chandra} observations and EPIC count\,s$^{-1}$ for the {\sl XMM-Newton} observations.
Because of the different efficiencies the psd$\sb{\rm n}$'s are expected to generally differ by a factor of
$\sim$25$\pm$5 for the same luminosity level given the energy spectrum described in section 2.1.
         
\begin{figure}[t]
\psfig{figure=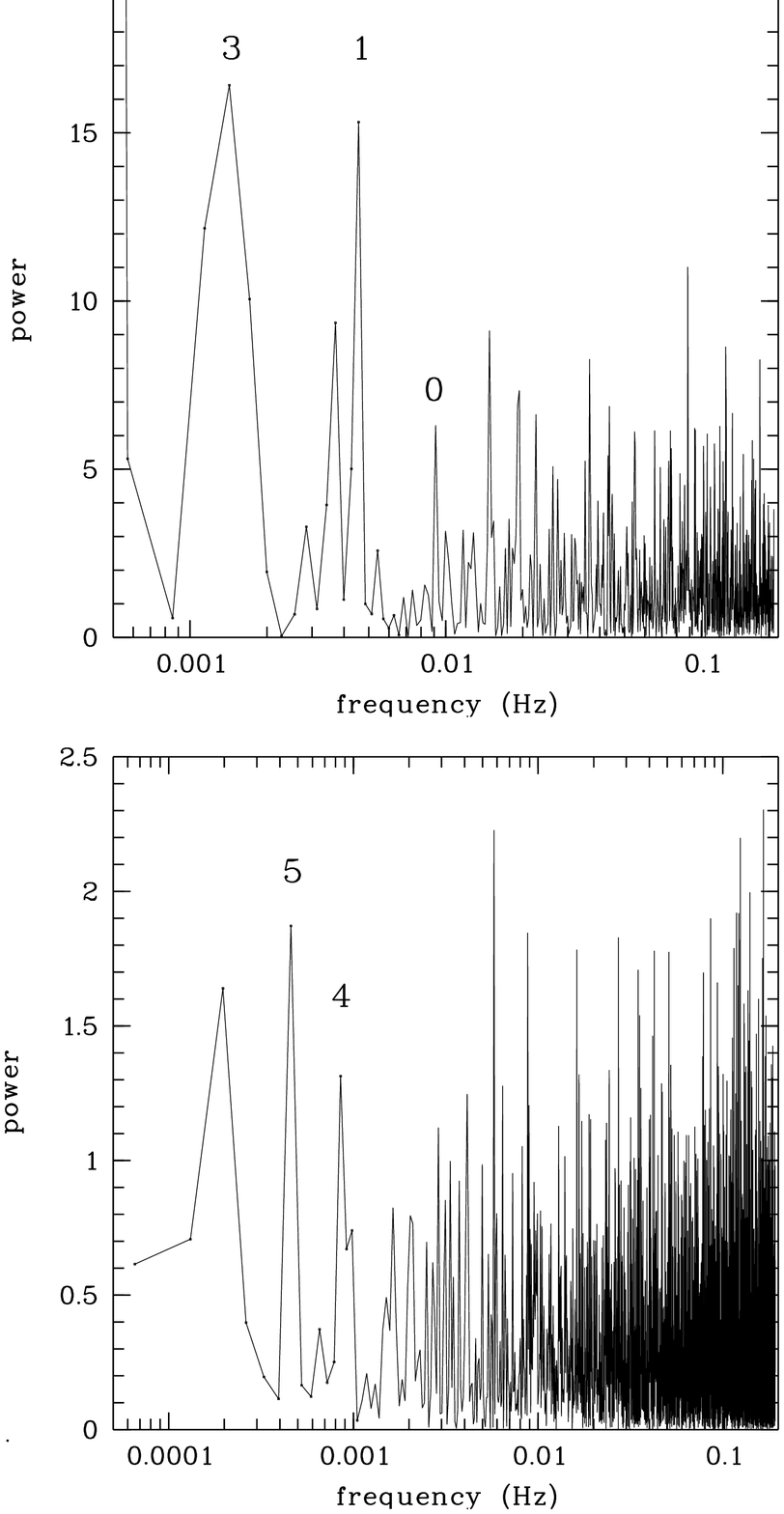,width=8cm,angle=0,%
bbllx=70pt,bblly=18pt,bburx=530pt,bbury=820pt,clip=}
%\vspace*{-0.56cm}%
\caption{Power density spectra of the October 3, 2002 flare (upper panel, Fig. 3a) 
and its precursor section (lower panel, Fig. 3b); {\sl XMM-Newton} measurement.
\vspace*{-0.4cm}
}
\label{fig3}
\end{figure}

\begin{figure}[t]
\psfig{figure=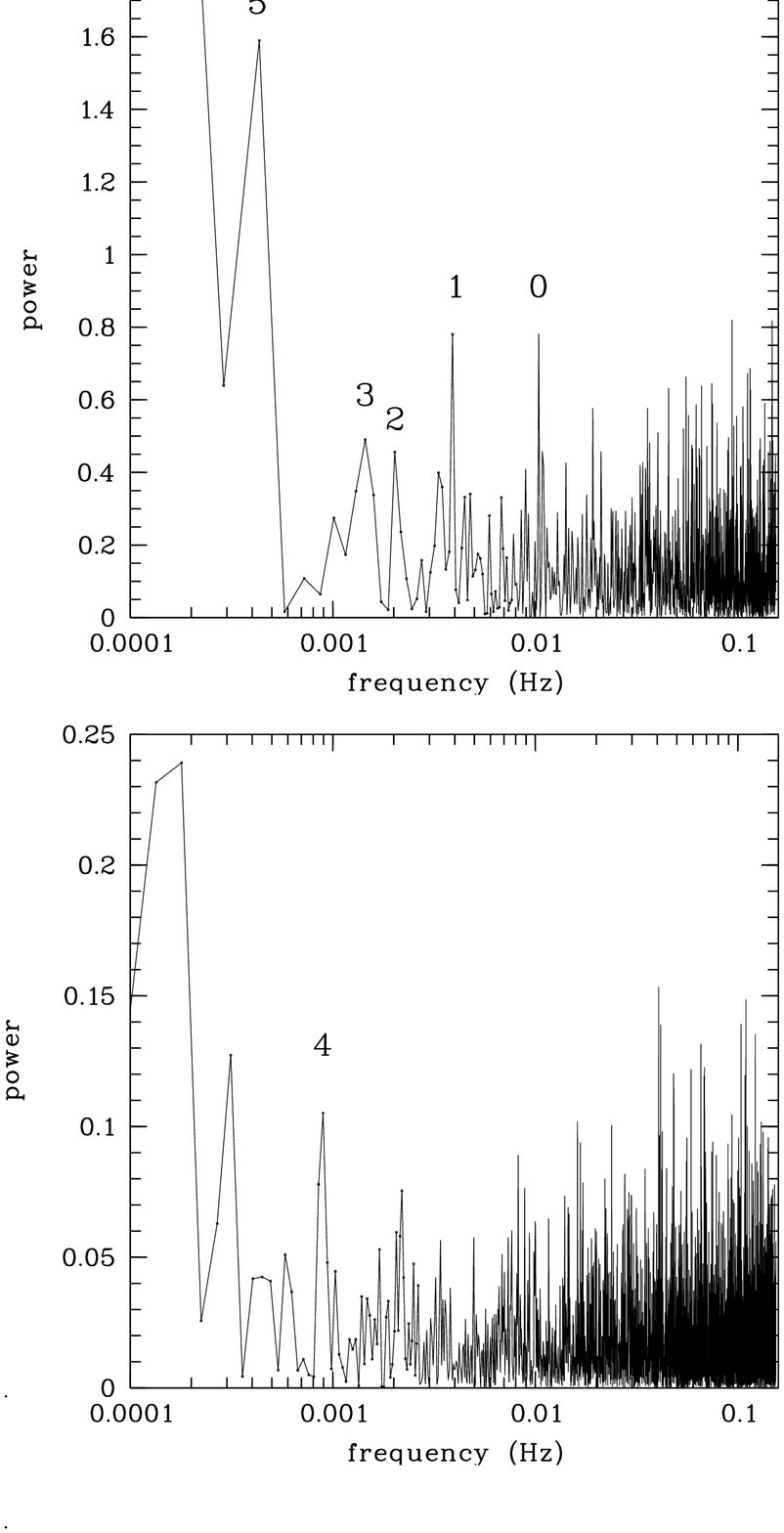,width=8cm,angle=0,%
bbllx=70pt,bblly=18pt,bburx=530pt,bbury=820pt,clip=}
%\vspace*{-0.56cm}%
\caption{Power density spectrum of the October 26, 2000 flare (upper panel, Fig. 4a)
and its precursor section (lower panel, Fig. 4b); {\sl Chandra} measurement.
\vspace*{-0.4cm}
}
\label{fig4}
\end{figure}

\begin{figure}[ht!]
\psfig{figure=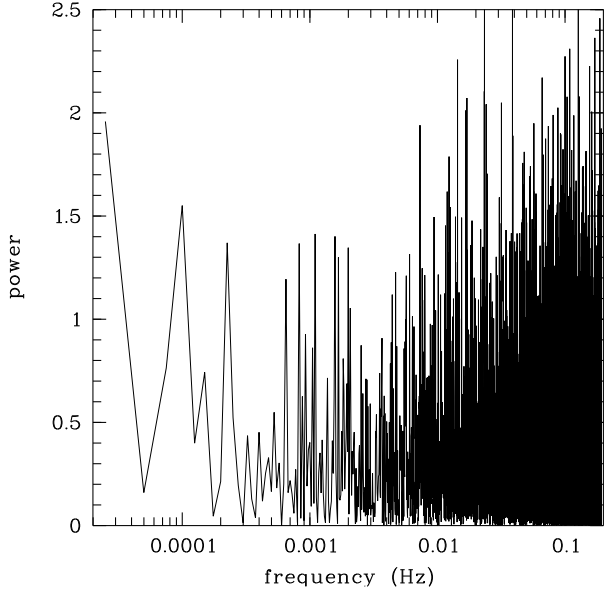,width=8cm,angle=0,%
bbllx=70pt,bblly=424pt,bburx=472pt,bbury=820pt,clip=}
%\vspace*{-0.56cm}%
\caption{Power density spectrum of the February 26, 2002 quiescent state; 
{\sl XMM-Newton} measurement.
\vspace*{-0.4cm}
}
\label{fig5}
\end{figure}

\begin{figure}[ht!]
\psfig{figure=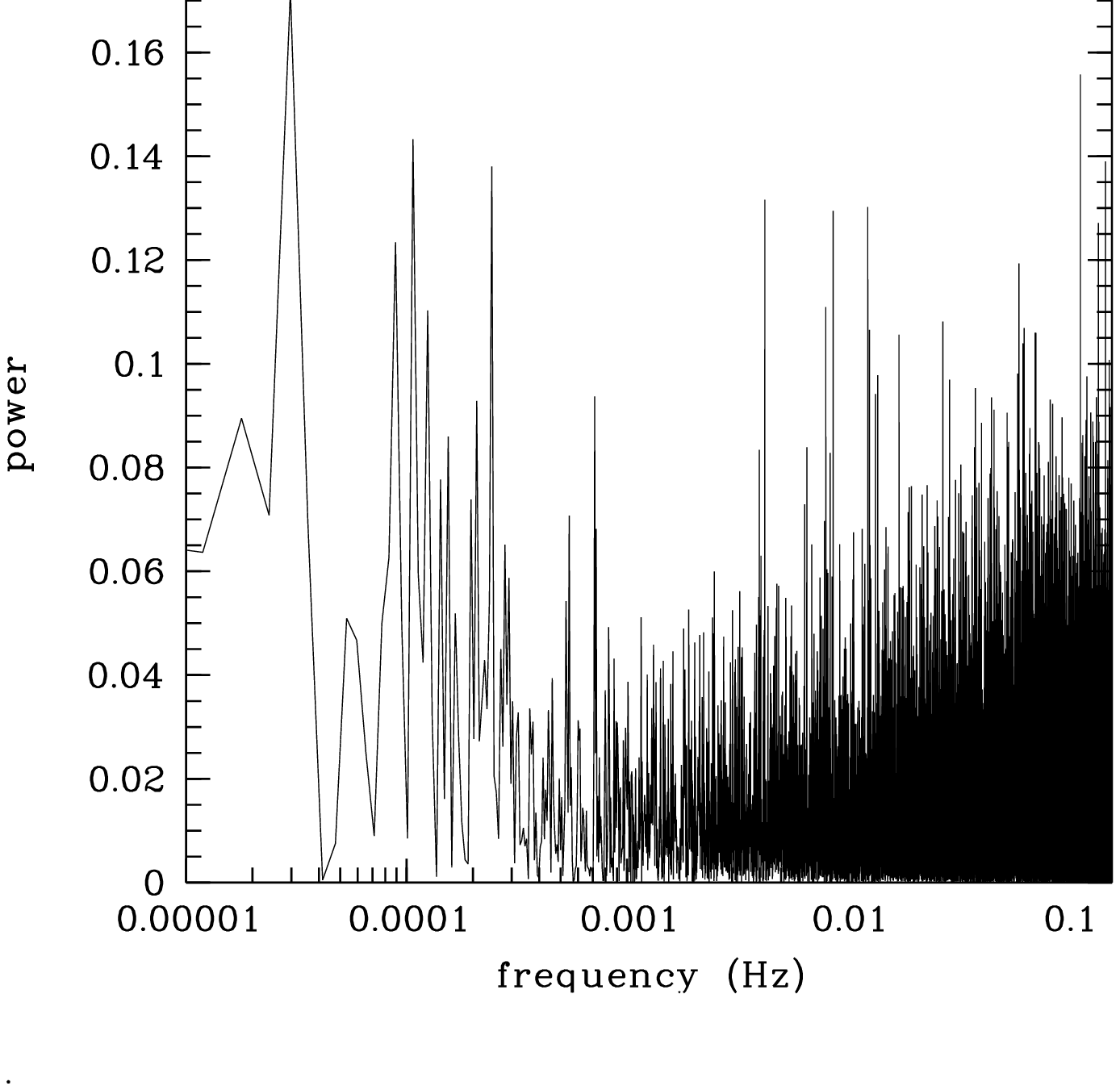,width=8cm,angle=0,%
bbllx=70pt,bblly=424pt,bburx=472pt,bbury=820pt,clip=}
%\vspace*{-0.56cm}%
\caption{Power density spectrum of the May 25, 2002 quiescent state; {\sl Chandra} measurement.
\vspace*{-0.4cm}
}
\label{fig6}
\end{figure}

\subsection {Frequencies}

In general the six PDS look very much alike. There is a high frequency component which is more or less 
rapidly growing with increasing frequency, which reflects the noise introduced by the 
low counting statistics. We note that the smallest possible binning has been chosen so that we deal 
with just no or one count per bin. At low frequencies the PDS are dominated by a few high peaks, 
which summarize the mean shape of the light curves, averaged over periods which are just a few times shorter 
than the whole track. Clearly separated from these two regimes is   
a  low noise mid-frequency section between $\sim$0.7\,mHz and $\sim$7\,mHz 
for the flare observations (Fig.~\ref{fig3}a and \ref{fig4}a),
 and 0.7\,mHz and 2.5\,mHz for the flare precursor sections (Figs.~\ref{fig3}b and \ref{fig4}b). 
In this frequency band we see two distinct PDS peaks 
at 1.426\,mHz  and 4.562\,mHz (Fig.~\ref{fig3}a; labels 3 and 1, respectively; {\sl XMM-Newton} flare) 
and similarly at 1.445\,mHz and at 3.902\,mHz (Fig.~\ref{fig4}a; labels 3 and 1, respectively; {\sl Chandra} flare), 
standing well above the noise level in the corresponding frequency bands. In the {\sl Chandra} flare 
we see a third peak at 10.41\,mHz (label 0), which, however, 
is located at the beginning of the climbing high frequency noise section (Fig.~\ref{fig4}a). The 
{\sl XMM-Newton} flare PDS shows a similar close-by peak at 9.12\,mHz (label 0), 
although at a fairly low power level. Like in the {\sl Chandra} case 
this frequency is at the high frequency edge of the low-noise 
band. The existence of this pair of frequencies might be doubtful but they should be noted 
basically because they seem to appear in two different observations. 
These are the only pairs in the flare section.  
Absent in the {\sl XMM-Newton} PDS but fairly prominent in the {\sl Chandra} PDS (Fig.~\ref{fig4}a)
is a peak at $\sim$2\,mHz (label 2). \\

The PDS of the data preceding the flare also show distinct peaks in the low noise regime at 
0.853\,mHz (Fig.~\ref{fig3}b, label 4; {\sl XMM-Newton}) 
and 0.895\,mHz (Fig.~\ref{fig4}b, label 4; {\sl Chandra}); 
a further peak appears in the {\sl XMM-Newton} PDS at 0.459\,mHz (label 5).  
The power at this frequency is so high that the equivalent period of 2178\,s can be identified by peaks 
in the light curve (c.f. arrows in Fig.~\ref{fig1}a). A PDS peak associated with or close to 0.459\,mHz is clearly 
absent in the {\sl Chandra} flare precursor observation but there is a peak at 0.434\,mHz 
in the {\sl Chandra} flare (label 5). \\

The PDS is a function of discrete frequencies given by f$\sb{n}$ = ${{n}\over{T}}$ with 
1$\le n \le int({{T\over{2~\Delta t}}}$) and $\Delta$t the binning size. 
Therefore, there is in principle a  systematic relative frequency uncertainty 
possible of $({{\Delta\rm f}\over{\rm f}})\sb{\rm n}$ = $\pm({{1}\over{\rm n -1}})$.
In Table~\ref{freqlog}, we summarize the results including frequency $f$, 
wave number $n$, period $P$, 
power spectral density psd and a label ID, which has the same value for 
frequencies close to each other. The same ID's are shown as labels in the PDS graphs. 
We find in the {\sl XMM-Newton} and the {\sl Chandra} observations 
five groups of periods, each of which is a pair with one 
member from {\sl XMM-Newton} and one from {\sl Chandra}. 
The periods of the members of each pair are almost identical, i.e. 
 110/96\,s (label 0), 219/256\,s (1), 701/692\,s (3),  1173/1117\,s (4), 2178/2307\,s (5).
 They appear to be not exactly identical but conceding the maximal 
possible uncertainty they are consistent with each other. This coincidence strongly supports their 
existence and suggests that each pair represents the same process. 

\begin{table}
   \caption[]{Compilation of outstanding frequencies. $\sp{1)}$ The psd is given in units of 
              (count\,s$^{-1}$)$\sp 2$ Hz$^{-1}$. For {\sl XMM EPIC} count rates and for {\sl Chandra}
              ACIS-I, count rates  are quoted. na: not applicable, i.e., not covered because the 
              flare didn't last long enough. The lower section of the table contains the frequencies 
              of high psd observed in the two infrared flares by Genzel et al. (\cite{Ge2003}). 
              The data have been read off their figure~\ref{fig2}c.  
               }
    \label{freqlog}
\begin{flushleft}
\begin{tabular}{lccccc}
            \noalign{\smallskip}
\hline
State/& & & & &     \\  
Instrum.&f&n&P&psd&ID\\
        &(mHz)& & (s) & $\sp{1)}$& \\
\hline
\noalign{\smallskip}
\multicolumn{6}{c}{flare} \\
\noalign{\smallskip}
% & & & & &     \\
XMM    &9.123 &32&110&6.3&0 \\
Chandra&10.41 &72&96&0.78&0 \\
\noalign{\smallskip}
% & & & & &     \\
XMM    &4.562 &16&219&15.3&1 \\
Chandra&3.902 &27&256&0.78&1 \\
\noalign{\smallskip}
% & & & & &     \\
XMM    &- &-&-&-&- \\
Chandra&2.023&14&494&0.46&2 \\
\noalign{\smallskip}
% & & & & &     \\
XMM    &1.426 &5&701&16.4&3\\
Chandra&1.445 &10&692&0.49&3 \\
\noalign{\smallskip}
XMM&na &na&na&na&na \\
Chandra&0.434 &3&2307&1.6&5\\
\hline
\noalign{\smallskip}
% & & & & &     \\
\multicolumn{6}{c}{precursor}  \\
\noalign{\smallskip}
% & & & & &     \\
XMM    &0.853&13&1173&1.3&4\\
Chandra&0.895&20&1117&0.11&4\\
\noalign{\smallskip}
% & & & & &     \\
XMM    &0.459&7&2178&1.9&5\\
Chandra&-&-&-&-&- \\
\hline
\hline
\noalign{\smallskip}
\multicolumn{6}{c}{IR flare}   \\
\noalign{\smallskip}
% & & & & &     \\
IR/16    &4.76&24&214&-&1\\
% & & & & &     \\
IR/15    &2.0&6&498&-&2\\
% & & & & &     \\
IR/16  &1.39&7&733&-&3\\
% & & & & &     \\
IR/15    &1.0&5&996&-&4\\
IR/16    &0.99&5&1026&-&4\\
\noalign{\smallskip}
\hline
\end{tabular}
\end{flushleft}
\end{table}

Genzel et al. (\cite{Ge2003}) have published the discovery of a 16.8$\pm$2\,min 
period in the two infrared flares observed 
on June 15 and June 16, 2003. A look at the two published PDS shows 
that there are more peaks, which we have read off from 
their figure~\ref{fig2}c and added to Table~\ref{freqlog} (IR/15, IR/16). 
Except the peak at 321\,s and wave number n = 16, appearing in the 
June 16 flare, 
 each frequency has a close-by counterpart in the {\sl XMM-Newton} and/or 
{\sl Chandra} PDS adding another finding to groups 1, 2, 3 and 4. Of course, the psd at 214\,s 
in IR/16 is fairly low but it shows up as a local maximum. There is no peak at the frequencies of group 5 
in the infrared observations, which however have been too short to search for it. Obviously, the infrared 
frequencies are fully consistent with the X-ray frequencies.   \\ 

We note that at none of the frequencies selected from the flares or precursors a prominent  
psd is evident in the quiescent level observations. The PDS of the quiescent level observations are 
shown in Fig.~\ref{fig5} ({\sl XMM-Newton}) and Fig.~\ref{fig6} ({\sl Chandra}). 
The {\sl XMM-Newton} PDS shows a set 
of six prominent PDS peaks between 0.65\,mHz and $\sim$2\,mHz 
centered on $\sim$ 1\,mHz. 
This frequency band embraces 
the 0.853/0.895\,mHz peaks of both the {\sl XMM-Newton} and {\sl Chandra} flare precursors. 
The psd's at the six frequencies vary between 1.05 and 1.42 (EPIC count\,s$^{-1}$)$\sp 2$ Hz$^{-1}$ 
which is close to what is observed as power for the 1173\,s period
 for the flare precursor (Table~\ref{freqlog}). 
A psd peak at the period of 2178\,s is not present in the quiescent observations. 
The PDS of the {\sl Chandra} quiescent level observations doesn't show 
prominent peaks around one mHz (Fig.~\ref{fig6}). 
But extrapolating  the {\sl XMM-Newton} psd of $\sim$1.2 (EPIC count\,s$^{-1}$)$\sp 2$ Hz$\sp{-1}$ 
the equivalent {\sl Chandra} psd is expected to be close to the noise level. 

The lowest frequencies accessible for this investigation are covered 
by the {\it Chandra} May 25, 2002 observation, with no bright flare though 
(Figure~\ref{fig6}). Starting at $\sim$0.1\,mHz the psd drops 
rapidly with frequency reaching the noise level minimum at $\sim$0.4\,mHz. The region 
between 0.18\,mHz and 0.35\,mHz seems to show excess power peaking at $\sim$ 0.24\,mHz or 4100\,s. 
In this band the {\sl XMM-Newton} PDS is very similar to the {\it Chandra} PDS with 
 fairly high peaks at 0.225\,mHz and 0.1\,mHz, but details like in the {\it Chandra} 
observation are not resolved because of the shorter exposure (Figure~\ref{fig5}). 
This PSD structure is probably not related to the occurrence rate
of flares, which has been estimated to 1.2$\pm$0.6 per day  by Baganoff (\cite{B2003}), which
is a factor of about 17 lower in frequency.  
Excluding 
the  0.18\,mHz -- 0.35\,mHz band the PDS, with 20 single psd values remaining,  
can be fitted by a power law with an index of roughly $-$2.3.   
We stress that the statistics are poor and we don't claim the 
existence of a QPO but the excess between 0.18\,mHz and  0.35\,mHz looks like a QPO structure. 
QPO's (quasi-periodic oscillations), which show up 
as a band of increased psd above an underlying, much broader PDS, which has the shape of a power law, have been 
observed in a number of galactic binaries, containing a compact object like a neutron star or black hole. The 
QPO's are considered to be created in associated accretion disks (e.g., van der Klis, \cite{vdK2000}). As  
Nowak \& Lehr (\cite{NL1998}) and others have pointed out, QPOs arising from supermassive black holes in AGN 
are expected to show periods between three hours and years, if a 0.1 s period is typical for a galactic black hole 
and if periods scale with mass. 
Such long periods are very difficult to measure for current missions. But for the Galactic Center black hole with 
its fairly low mass QPO periods might be shorter. In fact, if we scale the 300 Hz, which is the highest QPO frequency 
observed in the 7 M$_\odot$ black hole microquasar GRO J1655--40 (Remillard et al. \cite{RMC1999}, Orosz \&\ Bailyn \cite{OB1997})  
we expect a QPO around 0.6 mHz for the Galactic Center black hole, which is well within the long duration 
{\it Chandra} and {\it XMM-Newton} measurements.

\subsection{Statistical significance}

To quantify the significance of a measured psd value we have
approximated the rather erratic flare light curves by some smooth
functions which reproduce very well the general shape but do not introduce
high frequency variations in the regime which contains the
frequencies to be investigated. We are particularly interested
in the frequencies of group 0 and group 1. The lower frequencies
may be too much biased by red noise for which we don't have a model.
The model light curve
of the {\it{XMM-Newton}} flare consists of three exponentials, one for the
rise, one for the top and the third for the decay. The model light curve
for the {\it{Chandra}} flare is set up by a four piece polygon.
Each model light curve is normalized to the total number of measured
counts. The model light curves are binned with the same sampling time
as the measured data, and the model counts per bin are subject to
Poissonian statistics. For each flare Z = 3$\times$10$\sp 5$ light curves
have been produced, Fourier analysed and the PDS calculated.

The significance of a power peak is defined via a confidence level S,
which gives the
probability that the detected power psd$\sb{\rm{n, detect}}$ is not produced by
the noise process, i.e. 
1 - S = N$\sb{\rm{trials}}\times$W(psd$\sb{\rm n}\ge$ psd$\sb{\rm{n, detect}})$
(van der Klis, 1989).  
W is the ratio of the number
of powers exceeding psd$\sb{\rm{n, detect}}$ and Z. N$\sb{\rm{trials}}$ is the
number of frequencies for which the PDS has been measured, which is
674 and 1066 for the {\it{XMM-Newton}} and {\it{Chandra}} flare,
respectively. We stress that we have not made use of a fixed form for W like a
$\chi\sp 2$ or Gaussian distribution, but we built W by Monte Carlo
simulations (c.f. van der Klis, 1989).
The simulations show that there are some significant peaks, i.e.
for group 0: S(96 s) = 0.989 whereas the psd at 110 s of the
{\it{XMM-Newton}} flare is consistent with a noise created peak;
for group 1: S(219 s) = 0.993 and S(256 s) = 0.947, which means
that  signals with the latter periods quite likely exist.
There is just one more pair of close-by frequencies each of which shows a
significant psd, which are at a period of 11.31 s with S(11.31 s) = 0.634 in the
{\it{XMM-Newton}} flare and of 10.68 s with S(10.68 s) = 0.968 in the
{\it{Chandra}} flare. This very short period should be kept in mind for
future studies.

We also tried quite a few different model light curves but they do not change
significantly the values for
S given before.

\section{Discussion}

Five characteristic frequency groups have been discovered in the X-ray flares  
with their existence being supported by the 
fact that they have been found in more than one observation. 
We try to identify each of these frequencies with one of the oscillation modes 
which could occur in accretion disks surrounding black holes. 
As Nowak \& Lehr (\cite{NL1998}) for instance point out there are four cyclic modes which could  give rise 
to periodic or quasi-periodic changes in accretion disks. These modes are based on gravitation  
but it is an open question whether they would create changes of the light output at the very same frequencies. 
However, Nowak \& Lehr (\cite{NL1998}) point out that the mode frequencies 
predominantly depend upon fundamental 
gravitational frequencies and are not strongly affected by hydrodynamic processes for thin disks. 

\subsection{Characteristic accretion disk frequencies}   
 
There are four cyclic gravitational modes associated with black hole accretion disks 
(Nowak \& Lehr \cite{NL1998}, equations \ref{eq:1} to \ref{eq:3} and Merloni et al. \cite{MVSB1999}, 
equation \ref{eq:4}),  
which are the Kepler frequency ($\Omega\sb{\rm K}$), 
the disk perturbation frequencies in vertical and radial 
direction called vertical ($\Omega\sb{\rm V}$) and 
radial ($\Omega\sb{\rm R}$) epicyclic frequency and the Lense-Thirring 
precession frequency ($\Omega\sb{\rm{LT}}$). Each  frequency 
depends on the central mass M, 
the angular momentum $a$ and the radial distance $r$ from the center. 
Equations \ref{eq:1} to \ref{eq:4} show the relations, for which the 
standard notation of c=G=1 is used. Physical length scales 
 are in units of GM/c$\sp 2$ and angular frequencies 
$\Omega$ are in 
units of c$\sp 3$/GM. $r$ = 1 is defined as the gravitational radius r$\sb{\rm g}$. 

\begin{equation}\label{eq:1}
$$\Omega\sb{\rm K} = (r\sp{3/2} +  a)\sp{-1}$$
\end{equation}
\begin{equation}\label{eq:2}
$$\Omega\sb{\rm V}\sp 2 = \Omega\sb{\rm K}\sp 2 ~ (1 - {{4 a}\over{ r\sp{3/2}}} + {3 a\sp2\over{ r\sp 2}}) $$
\end{equation}
\begin{equation}\label{eq:3}
$$\Omega\sb{\rm R}\sp 2 = \Omega\sb{\rm K}\sp 2 ~ (1 - {6\over{r}} + {{8 a}\over{r\sp{3/2}}} - {3 a\sp2\over{ r\sp 2}}) $$
\end{equation}
\begin{equation}\label{eq:4}
$$\Omega\sb{\rm{LT}} = \Omega\sb{\rm K} - \Omega\sb{\rm V}$$.
\end{equation}

\subsection{Arrangement of frequencies and modes}

Equations \ref{eq:1} to \ref{eq:4} show that 
 $\Omega\sb{\rm{K}}$ decreases monotonically 
with increasing $r$ and, on the contrary, for $a$ fixed, both  
 $\Omega\sb{\rm{V}}$ and $\Omega\sb{\rm{R}}$ initially rise with $r$, go through a maximum and then 
decrease with increasing $r$. $\Omega\sb{\rm{V}}$ reaches its maximum at a radial distance denoted by    
 $r$ = r$\sb{\rm{Vmax}}$ and $\Omega\sb{\rm{R}}$ takes its maximum value at a 
distance denoted by $r$ = r$\sb{\rm{Rmax}}$.   
For any $r$ and $a$ fixed 
$\Omega\sb{\rm{K}}$ $>$ $\Omega\sb{\rm{V}}$ $>$ $\Omega\sb{\rm{R}}$. 
This means that the shortest period is 
to be assigned to $\Omega\sb{\rm{K}}$  
at the smallest $r$ possible which is the innermost 
marginally stable orbit at $r$ = r$\sb{\rm{ms}}$ defined by $\Omega\sb{\rm R}$ = 0. 
For each of the other frequencies there is 
a choice between $\Omega\sb{\rm{V}}( r$ = r$\sb{\rm{ms}})$, 
$\Omega\sb{\rm{V}}( r$ = r$\sb{\rm{Vmax}})$, 
$\Omega\sb{\rm{V}}(r$ = r$\sb{\rm{Rmax}})$, $\Omega\sb{\rm{R}}(r$ = r$\sb{\rm{Vmax}})$ and 
$\Omega\sb{\rm{R}}(r$ = r$\sb{\rm{Rmax}})$. $\Omega\sb{\rm{R}}(r$ = r$\sb{\rm{ms}})$
 is excluded because 
$\Omega\sb{\rm{R}}($r = r$\sb{\rm{ms}})$ = 0 per definition. Whether  
$\Omega\sb{\rm{LT}}$ could be generated at $r >$ r$\sb{\rm{ms}}$ is hard to tell, but 
$\Omega\sb{\rm{LT}}( r$ = r$\sb{\rm{ms}})$ could approach $\Omega\sb{\rm{K}}( r$ = r$\sb{\rm{ms}})$ 
or could even be lower than the frequency of any other mode, depending on $a$.  
We have chosen three radial fix points, i.e. $r$ = r$\sb{\rm{ms}}$, $r$ = r$\sb{\rm{Vmax}}$ and 
r = r$\sb{\rm{Rmax}}$. In principle there is no preference for any $r$ except  for $r$ = r$\sb{\rm{ms}}$ but 
we note that $\Omega\sb{\rm{V}}$ and $\Omega\sb{\rm{R}}$
 do not vary a great deal with $r$ over a reasonable range of $r$ centered on r$\sb{\rm{Vmax}}$ or 
r$\sb{\rm{Rmax}}$, respectively. In the following we use periods, 
i.e.  P$\sb{\rm{LT}}$, P$\sb{\rm K}$, P$\sb{\rm V}$ and P$\sb{\rm R}$ instead
of angular frequencies $\Omega$. 
Each observed period  is now assigned to one mode at either
  r$\sb{\rm{ms}}$, r$\sb{\rm{Vmax}}$ or r$\sb{\rm{Rmax}}$ 
each resulting in a relation between M$\sb{\rm{BH}}$ and $a$. If these assignments cover all 
observed periods within their measurement uncertainties, a common value of 
M$\sb{\rm{BH}}$ and $a$ should emerge. 
Group 0 contains the shortest period of 110\,s; if assigned to P$\sb{\rm{K}}($r = r$\sb{\rm{rms}})$ an upper 
limit for M$\sb{\rm{BH}} < $ 1.78$\times$10$\sp 6$M$_\odot$ exists for $a \rightarrow$ 1, which is in 
conflict with the lowest value of 2.2$\times$10$\sp 6$M$_\odot$ 
published by Sch\"odel et al. (\cite{S2002}). Therefore this ascription is excluded.
The next longer period of 219\,s implies an upper limit of 
M$\sb{\rm{BH}} < $ 3.55$\times$10$\sp 6$M$_\odot$ if assigned to P$\sb{\rm{K}}($r = r$\sb{\rm{rms}})$. 
This is  within the range of values for M$\sb{\rm{BH}}$ reported in the literature 
and suggests the assignment to Keplerian motion along the marginally stable orbit. 

The only solution of equations \ref{eq:1} to \ref{eq:4} with consistent values for M$\sb{\rm{BH}}$ and 
$a$ is obtained for the following assignation of period and mode:\hfill\break 
group 1, 219\,s = P$\sb{\rm{K}}(r$=r$\sb{\rm{ms}})$ [K]; \hfill\break
group 3, 692\,s = P$\sb{\rm{V}}(r$=r$\sb{\rm{ms}})$ [V]; \hfill\break
group 4, 1117\,s = P$\sb{\rm{R}}(r$=r$\sb{\rm{Rmax}})$ [R].  \hfill\break
With this set equation \ref{eq:4} predicts P$\sb{\rm{LT}} \approx$ 320\,s. Such a period shows up with increased power 
in the June 16 infrared flare (321\,s, Genzel et al. \cite{Ge2003}, c.f. section 2.3). 
The four relations for M$\sb{\rm{BH}}$ versus $(1 - a)$ following from 
equations \ref{eq:1} to \ref{eq:4} are drawn in figure~\ref{fig7}, which shows an almost perfect, single  intersection
 of the four relations. 
The wide error polygon of figure~\ref{fig7} indicates the maximal range for 
M$\sb{\rm{BH}}$ and (1 $-~a$) without the 321\,s period and 
the inner polygon covers the constraints set by the 321\,s period.
This solution does not suggest a specific radius like r$\sb{\rm{ms}}$,
 r$\sb{\rm{Vmax}}$ or r$\sb{\rm{Rmax}}$
 for the remaining period of $\sim$500 s, but 
it is consistent with a Kepler period expected close to $r$=r$\sb{\rm{Rmax}}$. 
The periods of group 5 around 2200--2300\,s, cannot arise from a region $r <$r$\sb{\rm{Rmax}}$. 
However, we point out that this period is very close to just twice the
 period of the radial epicyclic mode and might 
indicate a dominance of the amplitude of every second cycle. 
We note that the group 0 period of $\sim$100\,s is close to the first harmonic 
(m = 2) expected for the 
 'diskoseismology' C-mode 
(Nowak \& Lehr \cite{NL1998}).

The best-fit solution to the intersections is M$\sb{\rm{BH}}$ = 2.72$\sp{+0.12}\sb{-0.19}$ $\times$ 10$\sp 6$M$_\odot$
and $a$ = 0.9939$\sp{+0.0026}\sb{-0.0074}$. 
The errors quoted are not 1$\sigma$ errors but reflect the maximal possible
uncertainty given by the maximal uncertainties of the period determination. 
The radii involved are r$\sb{\rm{ms}}$ = 1.371  
and r$\sb{\rm{Rmax}}$ =2.507. The gravitational radius ($r$=1) amounts to 4.1$\times$10$\sp{11}$\,cm.    

\begin{figure}[t]
\psfig{figure=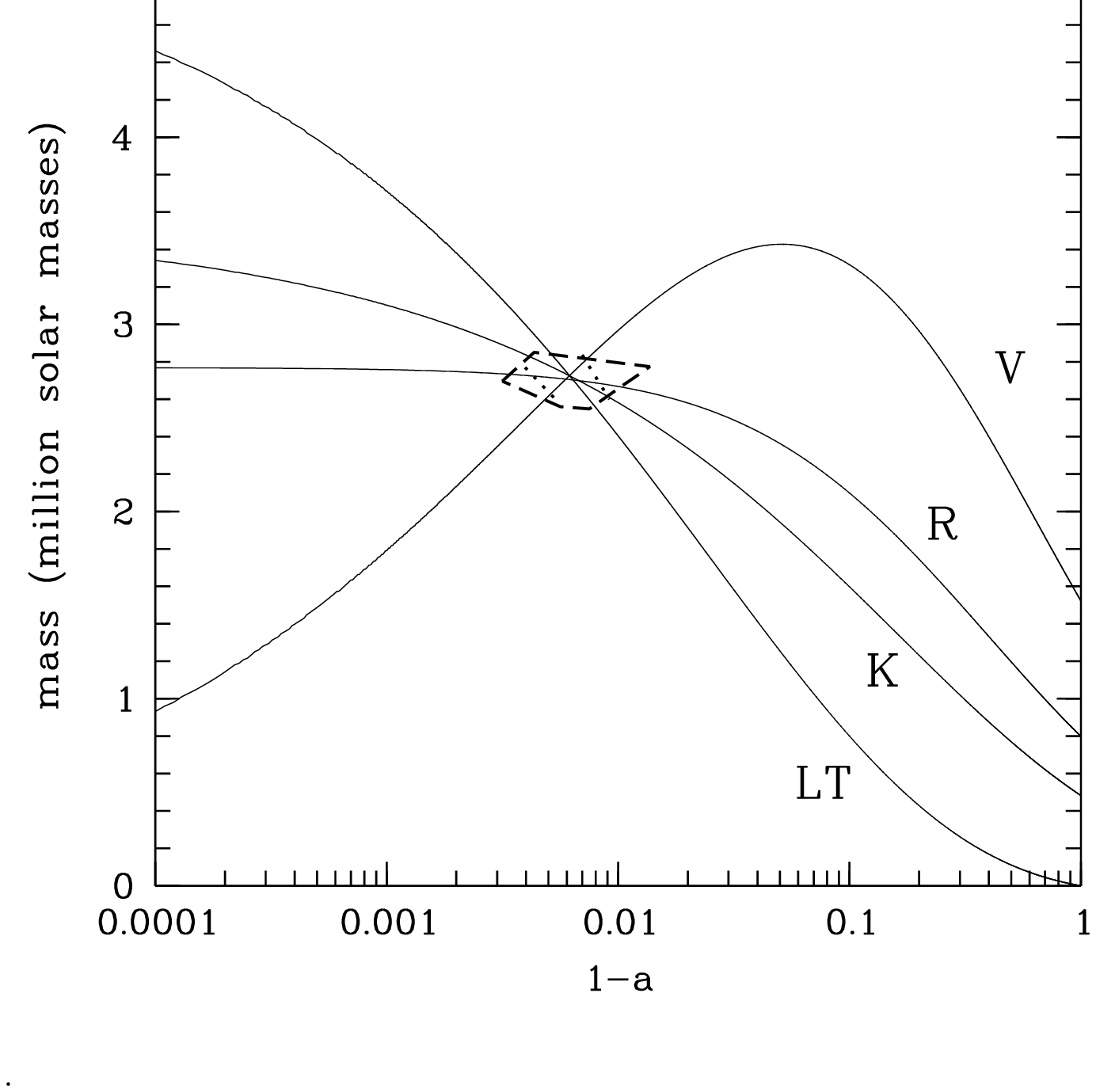,width=8cm,angle=0,%
bbllx=70pt,bblly=424pt,bburx=472pt,bbury=825pt,clip=}
%\vspace*{-0.56cm}%
\caption{Relation of M$\sb{\rm{BH}}$ versus (1 $-~ a$). Each curve represents one period assigned
to a specific gravitational mode, coded by a label  
described in the text. The error region is enclosed  by the dashed polygon. The dotted polygon 
is narrower because of the additional constraints imposed by the 321\,s period (LT).   
\vspace*{-0.4cm}
}
\label{fig7}
\end{figure} 

Independent measurements of the orbital motion of the star S0-2 (S2) around the GC black hole have
 resulted in M$\sb{\rm{BH}}$ = 3.7$\pm$1.5 $\times$ 10$\sp 6$M$_\odot$ (Sch\"odel et al. \cite{S2002}) 
and  M$\sb{\rm{BH}}$ = 4.07$\pm$0.62 $\times$ 10$\sp 6$M$_\odot$~(R$\sb 0/8 {\rm{kpc}})\sp 3$
 (Ghez et al. \cite{Gh2003}) with R$\sb 0$ the distance to the black hole.  
Our result is consistent with the Sch\"odel et al.~ measurement at their 
1~$\sigma$ level and it also agrees with Ghez et al.\,'s measurement at their 2$\sigma$ level and 
R$\sb 0$ = 8 kpc. Earlier Genzel et al. (\cite{GTK1996}) determined M$\sb{\rm{BH}}$ to 
2.5 - 3.0 $\times$ 10$\sp 6$M$_\odot$ from measurements of dark matter concentration in the central 
parsec of the Milky Way which appears to match our result best.   

\subsection{The high value of $a$}

The high value of $a$ $\approx$ 1 means that the emission from the inner parts of the accretion disk 
is quite close to the black hole. The event horizon is located at 
r$\sb{\rm H}$ = 1.112. With r$\sb{\rm{ms}}$ 
= 1.371 the distance to the marginally stable orbit is just 3.6 light seconds. Light from matter crossing the 
marginally stable orbit is therefore very likely to disappear in a fairly short time. 
Is the PDS peak at $\sim$11 s the characteristic time scale at which
the light is fading away?
The light curve of the {\it XMM-Newton} flare 
(figure~\ref{fig1}a) shows a deep, very short ($<$ 100 s) 
cut close to maximum light, which might be such an event. 
A similar event is also seen in the {\it Chandra} flare (figure~\ref{fig2}a). 
The arrows plotted in figure~\ref{fig1}a point to the maxima of the 2178\,s period, 
which seem somehow to synchronize with  
the deep cut in the light curve. According to our arrangement scheme 
the 2178\,s period is characteristic for the radial epicyclic mode and it looks as if matter is pushed across 
the marginally stable orbit. 

The high value of $a$ $\approx$ 1  and the proximity of the inner region to the event horizon also implies 
that fairly low ($<$1) general relativistic boost factors $g$ have a major impact on the emission because of 
increasing beaming (e.g. M\"uller \& Camenzind \cite{MC2004}). The radiation is not only shifted in frequency
but the observed flux density (F$\sb{\rm{o}}$) is significantly down-boosted compared to its rest-frame value
(F$\sb{\rm{rf}}$) with F$\sb{\rm{o}}$ =  $\overline{{g}\sp{3-\alpha}}\times\rm F\sb{\rm{rf}}$
 (Lind \& Blandford \cite{LB1985}) for a power law spectrum of energy index  $\alpha$, 
which is $-$1.5 for the {\it XMM-Newton} flare (Porquet et al. \cite{P2003}). 
$g^{3-\alpha}$  is to be averaged over an orbit, indicated by the overline, as $g$ varies along the orbit.  
For $a$ = 0.994, and a  
disk almost viewed edge-on at an inclination angle of $i$ = 70$\sp{\circ}$,  
 $g$ varies between 0.064 and 0.834 at $r$ = r$\sb{\rm{ms}}$ and  
$\overline{{g}\sp{3-\alpha}}$  = 0.056, 
or the radiation received by the observer is reduced by a factor of $\sim$18 compared to its rest frame 
value. For larger radii the reduction factor approaches unity, but at $r$ = r$\sb{\rm{Rmax}}$ = 2.5
 the radiation is still reduced by a factor of 1.7. 
The reduction becomes much more pronounced for disks viewed almost face-on. 
For $i$ = 15$\sp{\circ}$ 0.133  $\le$ $g \le$ 0.265 and $\overline{{g}\sp{3-\alpha}} <$ 0.001. 
The Lense-Thirring frequency and the Keplerian motion at $r$ = r$\sb{\rm{ms}}$ 
have become unobservable like any other frequency close to the marginal stable orbit. At $r$ = 2.5,  
$\overline{{g}\sp{3-\alpha}} <$ 0.038. Radiation from the innermost region is on average  
dimmed by a factor of some hundred, depending on the details of the radial brightness distribution. 
Because of the large reduction factors the observation of the inner regions of almost face-on accretion disks 
surrounding supermassive extragalactic black holes in AGN becomes progressively more difficult with 
increasing $a$ and/or decreasing $i$. Maybe that this effect contributes 
to some extent to the faintness of the GC black hole.

Although we have no suggestion which physical process gives rise 
to the flare we constrain the region from where it 
comes. Since we see the light modulated with frequencies associated 
with the marginally stable orbit out to the maximum frequency of the radial 
epicyclic mode and in between the bulk of the emission arises from a region 
of 1.37 $<~  r ~ <$ 2.5, i.e. the innermost region of the accretion disk.

%The over the subjected 
%to fairly low ($\ll$1) general relativistic boost factors g, i.e. the radiation is not only shifted in frequency 
%but the observed flux density (F$\sb{\rm{o}}$) is significantly down-boosted compared to its rest-frame value
%(F$\sb{\rm{rf}}$) with F$\sb{\rm{o}}$ =  $\overline{\rm{g}\sp{3-\alpha}}\cdot$F$\sb{\rm{rf}}$ (Lind \& Blandford, 
%1985) for a power law spectrum of energy index  $\alpha$, which is -1.5 for the {\it XMM-Newton} flare 
%(Porquet et al., 2003). $\rm g\sp{3-\alpha}$  is to be averaged over an orbit indicated by the overline.    
%

\section{Conclusion}

We have discovered five distinct periods in the power 
density spectrum of the {\it XMM-Newton} X-ray  
flare of October 3, 2002. One may wonder about the statistical
 significance, but these five periods, within their 
measurement uncertainty,  also appear in the power density spectrum
 of the October 26, 2000 X-ray flare observed by {\it Chandra}, with perhaps 
one additional period which, however, does not show up in the {\it XMM-Newton} flare. 
Further evidence for the existence of these periods comes from the June 16, 2003 infrared flare
 (Genzel et al. \cite{Ge2003}). The power density spectrum 
shows a clear increase for at least two of the periods and possibly 
for a third period; the remaining two periods 
were not accessible to the infrared observations. Each of the periods can be identified with one of the 
four characteristic gravitational modes in accretion disks, i.e. Lense-Thirring precession, Kepler motion, 
vertical and radial epicyclic oscillation, in such a way that a common value
 for the black hole mass M$\sb{\rm{BH}}$ 
and the angular momentum $a$ emerges, i.e., 
M$\sb{\rm{BH}}$ = 2.72$\sp{+0.12}\sb{-0.19}$ $\times$ 10$\sp 6$M$_\odot$
and $a$ = 0.9939$\sp{+0.0026}\sb{-0.0074}$. 
The available data on M$\sb{\rm{BH}}$ derived from studies of the orbital motion of the S0-2 star 
and the dark matter concentration in the center of the Milky Way 
are consistent with our result concerning M$\sb{\rm{BH}}$, which, by the way, 
is independent of the distance to the black hole.  
There are indications of a $\sim$100\,s periodicity in the 
X-ray data, both in {\it XMM-Newton} and {\it Chandra}, 
but right now, we are reluctant to accept it as firmly established. For further progress in the X-ray 
domain a flare even brighter than the {\it XMM-Newton} flare is required, 
and in the infrared band the temporal 
resolution should be improved to $\sim$10\,s. This is particularly important 
for simultaneous multi-wavelength observations. 

%
%Our findings demonstrate that at least one black hole exists with a fairly high value of a, which implies 
%that a very large fraction of the luminosity generated in the inner 
%part of the accretion disk (r$<$3) is not received by the observer because of the very small 
%($\ll$1) general relativistic boost factors. The overall dimming  
%depends of course on the radial brightness distribution produced in the rest frame. 
%If the distribution is highly peaked to smaller radii both a high value of a and a low inclination 
%angle of the accretion disk could significantly reduce their observed luminosity compared 
%to otherwise similar black holes but of low a and large inclination angle. 
%

\begin{acknowledgements}
B.A. likes to thank Wolfgang Brinkmann, MPE Garching, for numerous inspiring discussions and Andreas
M\"uller, LSW Heidelberg, for providing the relativistic boost factors.
We are grateful to Andrea Merloni, MPA Garching, for pointing out to us the correct formula for 
the Lense-Thirring precession frequency. 
D.P. is supported by a MPE fellowship.
\end{acknowledgements}

           \end{document}